\documentstyle[aps, preprint]{revtex}
\tightenlines

\def\v{\mbox{\bf v}}
\def\r{\mbox{\bf r}}
\def\k{\mbox{\bf k}}
\def\p{\mbox{\bf p}}

\def\be{\begin{equation}}
\def\ee{\end{equation}}
\begin{document}
\input epsf
\draft
\preprint{PURD-TH-99-08, astro-ph/9911218}
\date{November 1999}

\title{Short-scale gravitational instability in a disordered Bose gas}
\author{S. Khlebnikov}
\address{
Department of Physics, Purdue University, West Lafayette, IN 47907, USA
}
\maketitle

\begin{abstract}
We study numerically evolution of a self-gravitating nonequilibrium Bose gas 
contained in a fixed volume. We find that, even when the volume
is small enough to prevent collisionless instability, a compact 
phase-correlated object (a Bose star) can form, by gravity alone,
from a disordered initial state. We interpret this effect 
and the associated growth of the density contrast as consequences of a new type 
of gravitational instability---a nonlinear instability due to stimulated 
gravitational scattering of the bosons. Our results imply that formation of 
Bose stars, in regions that have fallen 
out of the Hubble expansion, may be a quite general phenomenon, not requiring 
a large preexisting correlation length or any short-range interactions.
\end{abstract}

\pacs{}

\section{Introduction}
Gravitational (Jeans \cite{Jeans}) instability is believed to be at the root of
all the large-scale structure that we observe in the universe. (For a review
on structure formation, see ref. \cite{Peebles}.) The instability is a direct 
consequence of gravity being an attractive force: particles will accrue on small 
clumps, and the clumps will get bigger. 
Jeans's original calculation was based on description of matter 
via classical hydrodynamics. This description applies only 
at spatial scales larger than the mean-free path of the particles
(which is defined here with respect to some interaction other than gravity).
Evolution of density perturbations with scales smaller than
the mean-free path can for many purposes be described by the collisionless 
Vlasov approximation---unless the scale is so small that this semiclassical 
method breaks down. In the collisionless approximation, a self-gravitating gas
is unstable with respect to sufficiently long-wave perturbations, 
and the instability is of the same type as Jeans's. 
(We will define later more precisely what ``the same type'' means.)

For a wide class of initial distributions, however,
the collisionless instability disappears for  short-scale perturbations.
Consider, for example, a homogeneous (on average) nonrelativistic gas of 
particles of mass $m$ in a box of fixed size, in the Newtonian approximation 
to gravity (in addition, we assume that the uniform component 
of density does not gravitate). 
We show in Appendix that, for the class of momentum distributions defined there,
the collisionless instability exists only for density perturbations with 
wavenumbers
\be
k < C k_J^2 / k_0 \equiv k_{*} \; ,
\label{kcond}
\ee
where $k_0$ is a typical momentum of the particles, and 
$C$ is a numerical constant depending on their momentum distribution.
(We use a system of units with $\hbar = 1$.)
The wavenumber $k_{J}$ is determined by the average density 
number $n_{\rm ave}$:
\be
k_{J}^{4} = 16\pi G m^{3} n_{\rm ave} \; ,
\label{kJ}
\ee
and will be called the Jeans wavenumber. One may wonder
if any gravitational instability exists for perturbations with
$k > k_{*}$, that is at shorter scales.

The condition (\ref{kcond}) suggests that, depending on the
relation between $k_{0}$ and $k_{J}$, a self-gravitating gas can be in one 
of two different regimes. When
\be
k_{J} \gg k_{0} \; ,
\label{ordered}
\ee
eq. (\ref{kcond}) shows that collisionless instability occurs for perturbations 
with all wavenumbers up to
$k \sim k_{0}$, at which point the semiclassical description of particles 
breaks down, and the Vlasov equation can no longer be used. In at 
least one case, however, one can show that a linear instability exists
even for $k > k_{0}$ (although of course the argument cannot be based on 
the Vlasov equation). This case is a perfectly ordered Bose gas, i.e. 
a gas in which all particles are Bose-Einstein condensed in the $k=0$ 
mode. Elementary excitations of this Bose-Einstein condensate are Bogoliubov's 
quasiparticles, and their dispersion law is known, from Bogoliubov's 
work \cite{Bogoliubov}, for arbitrary two-body potential. When the 
only interaction between particles is Newtonian gravity, the 
dispersion law (for a nonrelativistic gas) reads
\be
\omega(k) = (1/2m) \left[ k^4  - k_{J}^{4} \right]^{1/2} \; ,
\label{disp1}
\end{equation}
which means that the perfectly ordered Bose gas is gravitationally 
unstable for all $k<k_{J}$. This result had been 
reported previously in a number of papers \cite{Khlopov&al,Bianchi&al}.
(We have presented it here for a region that is out of the Hubble 
expansion; a treatment in the expanding universe is also available 
\cite{Bianchi&al,Nambu&Sasaki}.)

The perfectly ordered state is just one possible initial state of a 
Bose gas. A generic initial state is a patchwork of correlated 
domains, rather than one such domain. If we 
describe a nonrelativistic Bose gas by a complex field $\psi$, 
correlated regions are those in which the phase of $\psi$ is more or 
less constant in space. The typical size of correlated regions is 
$2\pi/k_{0}$. If such a region contains many particles, we can apply 
formula (\ref{disp1}) to it individually, but only for 
wavenumbers $k \gg k_{0}$: only these sample a more or less ordered 
background. Eq. (\ref{disp1})  then predicts an instability for $k$ in the range
$k_{0} \ll k < k_{J}$. At $k\sim k_{0}$, we expect this instability 
to cross over smoothly to the instability found in the Vlasov approximation.
Thus, we expect that a nonrelativistic 
Bose gas satisfying the condition (\ref{ordered}) is unstable with 
respect to density perturbations with all wavenumbers $k< k_{J}$
(although the theoretical description of the instability has to be switched
at $k\sim k_{0}$).

We can now formulate more precisely what we mean by saying that 
the gravitational instability in the perfectly ordered Bose gas is 
of the same type as 
the instability found by Jeans for matter described by classical 
hydrodynamics.
The similarity has two closely related aspects. 
First, in either 
case, the instability is {\em linear}, i.e. density perturbations 
with different wavenumbers do not have to interact with one another for it to occur. 
Second, the growth rate of the instability, for low-$k$ modes, scales 
as square root of the average number density:
\be
{\rm Im} \omega(0) \propto \sqrt{n_{\rm ave}} 
\label{sqrt}
\ee
(see e.g. eq. (\ref{disp1})). One can say that this square-root
scaling law defines a certain ``universality class'', to which the original Jeans
instability, the collisionless instability, and the instability of the 
ordered Bose gas all belong.

In this paper, we want to consider a nonrelativistic Bose gas 
in which the relation between $k_{0}$ and $k_{J}$ is 
reversed compared to (\ref{ordered}):
\be
k_{J} \ll k_{0} \; .
\label{disord}
\ee
Our motivation is simply that we do not know at present which type of 
initial state is more likely to be realized in the early universe.
We will call the case (\ref{disord}) {\em 
disordered} (because correlated domains are initially small), 
as opposed to the case (\ref{ordered}), 
which we will call {\em ordered} (initial correlated domains are large). 
According to eq. (\ref{kcond}), the collisionless instability will 
occur in a disordered gas only at large spatial scales. 
So, if there is a gravitational instability on shorter scales, it is
caused by gravitational scattering of the particles, rather than by collisionless 
effects. Our goal was to determine effects of gravitational scattering on the evolution 
of a disordered Bose gas.

We imagine the following sequence of events
in the early universe. A disordered Bose gas first becomes unstable with respect to 
large-scale density perturbations via the collisionless (Vlasov)
mechanism. As clumps formed by this mechanism grow denser, 
collisionless instability can develop on shorter scales. We expect 
that this process will give rise to clusters of matter, each cluster 
including a number of smaller clumps and some inter-clump gas. 
Gravitational scattering, which is a slower process, will operate 
inside those clusters, after they have already fallen out of the 
Hubble expansion. Accordingly, we considered
a disordered Bose gas in a box of fixed size. In addition, we used Newtonian gravity,
which we expect to be good for sufficiently short-scale perturbations.

Our results, obtained by numerically following the 
evolution of a random classical field, apply when the initial state has 
large (compared to unity) occupation numbers at low momenta and little
occupation at high momenta. These results are as follows. (i) 
Starting from a disordered state,
we have observed a well-defined crossover from collisionless 
instability to a slower growth of the density contrast. This slower 
growth can be viewed as a gravitational instability of a different 
type than the linear instabilities discussed above.
We confirm the existence of this new type of 
gravitational instability by running simulations in smaller boxes, so 
as to remove the collisionless instability altogether. In this case, any growth
of the density contrast will have to be attributed to some new type of 
instability. We indeed observe such a growth, and we interpret 
it as a result of gravitational scattering of the bosons. 
The scaling of the rate of the growth with $n_{\rm ave}$, in a small box, is quite
distinct from the square-root law (\ref{sqrt}). In this sense, the new 
instability is in a different ``universality class'' than the linear
instabilities. 
(ii) Clumps that result from this nonlinear instability are 
phase-coherent; in other words, the instability leads to formation of 
Bose stars. (By a Bose star we mean any coherent 
gravitationally bound configuration of a Bose field; ground states of
gravitating bosons were originally discussed in ref. \cite{RB}.)
It has been long considered possible that Bose stars will form as 
result of the linear instability in an ordered Bose gas 
\cite{Bianchi&al,Grasso} 
(although, to our knowledge, that had not been actually 
demonstrated). Formation of Bose stars in an initially disordered 
gas is a different effect. We attribute it to gravitational 
scattering in our case being {\em stimulated}: as an effect of Bose statistics, 
particles that scatter into regions of high density will prefer to end up in phase 
with particles that are already there. 

The rest of the paper is organized as follows. Sect. 2 contains 
formulation of the problem in terms of a random classical field.
In Sect. 3 we present numerical results. Our conclusions, and connections to 
earlier work, are given in Sect. 4. Some technical details concerning the 
collisionless instability are given in Appendix.

 \section{Setting up  a random classical field}
We consider a nonrelativistic Bose gas described by the following 
equations:
\begin{eqnarray}
i\frac{\partial \psi}{\partial t} & = & -\frac{\nabla^{2}}{2m} \psi + 
m\Phi \psi + g_{4} \psi^{\dagger} \psi^{2} \; ; \label{eqpsi} \\
\nabla^{2} \Phi & = & 4\pi G m (\psi^{\dagger} \psi - n_{\rm ave}) 
\; . \label{eqPhi}
\end{eqnarray}
Here $\psi$ is a complex Bose field, and $\Phi$ is the (real) 
gravitational potential. Number density of bosons is $n=\psi^{\dagger} 
\psi$, and its volume average is called $n_{\rm ave}$. For 
reasons given in the introduction, we assume it sufficient to consider 
the gas in a box of fixed size (rather than in an expanding universe) 
and to use the Newtonian approximation to gravity. 

All numerical results presented in the next section are for $g_{4}=0$,
i.e. for the case when gravity is the only interaction between the 
particles. The method that we use can be applied equally well
to the case $g_{4} \neq 0$. In this paper, however, we limit 
ourselves to only a cursory discussion (in the concluding section) 
of possible effects of short-range interactions.

The field $\psi$ is in principle a quantum operator but here we treat 
it as a random classical field. Usefulness of this classical 
approximation has been recently emphasized in connection with a 
number of problems in nonequilibrium statistics of Bose fields. 
Among these problems, the one closest to the problem at hand
is phase separation in a nonequilibrium Bose gas with a {\em local}
attractive interaction (and a repulsive interaction at larger 
densities) \cite{dew}. Some of the results that we obtain here, 
notably, the coherent nature of the clumps, are similar to those
for a local interaction, but due to the long-range 
nature of gravity there are also some differences. For example, for 
a short-range interaction, one can also identify ordered and disordered types
of initial states. Results of ref. \cite{dew} correspond to a disordered initial 
state. For neither type of state, however, we expect a large-distance crossover 
to collisionless (Vlasov) dynamics.

The system (\ref{eqpsi})--(\ref{eqPhi}) can be viewed as a 
modification, required to include  Newtonian gravity, of
the Gross-Pitaevskii (GP) equation familiar from the theory of 
laboratory Bose gases \cite{Pitaevskii}.
The system (\ref{eqpsi})--(\ref{eqPhi})
is universal in the same sense as the GP equation is: at low gas 
densities, the only parameter needed to characterize short-range
interactions is the scattering length, which is encoded in the 
coupling constant $g_{4}$.

The classical approximation is good when the field $\psi$ is large and 
its power spectrum is contained at small momenta:
speaking in terms of particles,
the occupation numbers in low-momentum modes are 
much larger than unity, while those in high-momentum modes are small and 
rapidly decrease with momentum. We stress, however, that when 
scattering is strong, as it may become at later stages of clumping,
the very notion of a particle, as an entity propagating for 
relatively large times without collisions, becomes ill-defined.
One of the useful features of the classical 
approximation is that it allows one to avoid dealing with that notion
altogether. In addition, provided the power spectrum of $\psi$ satisfies the 
necessary conditions, the classical approximation will apply equally 
well either in the ordered or in the disordered case. We also stress 
that, although this may sound paradoxical, the classical approximation does 
take into account effects of the quantum Bose statistics of the 
particles. Indeed, the very possibility for $\psi$ to become (almost) 
classical, in the limit of large occupation numbers, is one such effect.

For numerical simulations presented in this paper, 
we have chosen the initial power spectrum of 
$\psi$ in the following  simple form (cf.  ref. \cite{dew}):
\be
|\psi_{\k}|^{2} = A \exp(-k^{2}/k_{0}^{2}) \; ,
\label{pws}
\ee
where $\psi_{\k}$ are Fourier components of the field $\psi$:
\be
\psi(\r) = V^{-1/2}\sum_{\k} \psi_{\k} \exp(i\k\r) \; ,
\label{psik}
\ee
$V$ is the integration volume. 
For an ideal gas, $|\psi_{\k}|^{2} $ would be the conventionally 
defined occupation numbers; we will often use the same 
terminology for the interacting gas.
For the classical approximation to apply, 
it is sufficient that $A\gg 1$. Although the distribution (\ref{pws}) 
may look like an equilibrium Maxwell distribution, it is totally 
unrelated to the latter. The Maxwell distribution follows from the Bose
distribution in the limit when the occupation numbers are small. In our 
case, $A\gg 1$, and the occupation numbers (at low wavenumbers) 
are large. Thus, the spectrum (\ref{pws}) is highly nonthermal.
 Eq. (\ref{pws}) determines the magnitudes of
$\psi_{\k}$. The phases of $\psi_{\k}$ are chosen as uncorrelated 
random numbers. The initial number density contrast corresponding to 
these initial conditions is $\delta n/n \approx 1$.

Bose gases with highly nonthermal spectra can form in the early 
universe through various nonequilibrium processes, such as spinodal 
decomposition or postinflationary reheating. (Formation of nonthermal
spectra in the latter case is reviewed in \cite{eger}.) 
We do not pretend that the specific form
(\ref{pws}) is in any way realistic. We only use it as a 
representative of spectra satisfying the 
conditions of applicability of the classical approximation. We have 
also considered some different forms of the initial spectra and have 
convinced ourselves that these modifications do not alter the results in 
any significant way.

Finally, for comparison with the classical method used here, we review results 
obtained in the collisionless (Vlasov) approximation. Evolution of density 
perturbations with wavenumbers $k \ll k_{0}$ can be studied by 
considering particles as semiclassical wave packets. We stress 
that this 
semiclassical approximation for the particles is entirely different 
from the classical approximation for the field $\psi$. In particular, 
it does not require that the occupation numbers be large but requires 
that perturbations be slowly varying.  (It is the other way around for
the approach based on a random classical field.) 
When the semiclassical view of the particles applies (i.e. 
a perturbation is slowly varying), one can describe them with a 
classical distribution function $f(\r,\p;t)$, which is proportional to the Fourier 
transform, with respect to $\k$, of the quantum average of 
$\psi^{\dagger}_{\p-\k/2} \psi_{\p+\k/2}$. The distribution function,
together with the gravitational potential $\Phi$,  can be used 
to construct a collisionless approximation (which, if needed, can be 
augmented by a collision integral). The construction is entirely 
analogous to the derivation of the Vlasov equation for plasmas, see 
e.g. \cite{LP}. In the collisionless approximation, we find that
frequencies $\omega(\k)$ of small perturbations of a homogeneous state with a
distribution function $f_{0}(\p)$ are determined by the dispersion 
equation
\be
1 + \frac{4\pi G m^{2}}{k^{2}} \int \k {\partial f_{0} \over 
\partial \p} \frac{d^{3} p}{\k\v - \omega -i0} = 0 \; ,
\label{dispeq}
\ee
where $\v=\p/m$. When $kv$, for a typical $v$, is much smaller than $|\omega|$, 
the dispersion equation reduces to
\be
1 + 4\pi G m n_{\rm ave} /\omega^{2} = 0 \; .
\label{disp2}
\ee
Eq. (\ref{disp2})  shows that there is a linear instability with the growth rate
\be
\omega_{i} (0) = (4\pi G m n_{\rm ave})^{1/2} = k_{J}^{2}/2m \; .
\label{cless}
\ee
For a wide class of initial distribution functions, which is specified in 
Appendix, the collisionless
instability disappears at $k \geq k_{*}$, where $k_{*}$ is estimated 
as in the condition (\ref{kcond}). The precise value for $k_{*}$ is obtained 
by setting $\omega = 0$ in eq. (\ref{dispeq}) and solving for $k$. 
(This procedure is justified in the Appendix.)
For example, in an initial state with power spectrum (\ref{pws}) the
distribution function is
\be
f_{0}(\p) = \frac{A}{(2\pi)^{3}} \exp(-p^{2}/p_{0}^{2})  \; ,
\label{f0}
\ee
where $p_0 = k_0$ in the system of units with $\hbar = 1$.
In this case, collisionless instability occurs only for
\be
k <  k_{*} = (\xi / 2)^{1/2} k_{0} \; ,
\label{k*}
\ee
where $\xi = k_{J}^{4}/k_{0}^{4}$. Eq. 
(\ref{k*}) is equivalent to (\ref{kcond}) with $C=1/\sqrt{2}$.

\section{Numerical results}
For numerical work, it is convenient to rescale the fields of our 
model, so that we have to deal with a smaller number of parameters.
Define new fields $\chi$ and $\phi$ as
\begin{eqnarray}
\chi & = & (8\pi G m^{3})^{1/2} \psi \; , \label{chi} \\
\phi & = & 2 m^{2} \Phi \; . \label{phi}
\end{eqnarray}
Then, eqs. (\ref{eqpsi})--(\ref{eqPhi}) with $g_4=0$ take the form
\begin{eqnarray}
2mi\frac{\partial \chi}{\partial t} & = & -\nabla^{2} \chi + \phi \chi \; , 
\label{eqchi} \\
\nabla^{2} \phi & = &  \chi^{\dagger} \chi - \nu_{\rm ave} \; , \label{eqphi}
\end{eqnarray}
where $\nu_{\rm ave}$ is the average of $\nu  = \chi^{\dagger} \chi$
over space. Initial condition (\ref{pws}) takes the form
\be
|\chi_{\k}|^{2} = B \exp(-k^{2}/k_{0}^{2}) \; ,
\label{pwschi}
\ee
where $B = 8\pi G m^{3}A$, and $\chi_{\k}$ are related to $\chi(\r)$ in
the same way as $\psi_{\k}$ to $\psi(\r)$, see eq. (\ref{psik}). Finally, the 
expression (\ref{kJ}) for the Jeans wavenumber reduces to
\be
k_{J}^{4} = 2 \nu_{\rm ave} \; .
\label{kJtil}
\ee

All data presented in the figures were obtained by numerically integrating 
the system (\ref{chi})--(\ref{phi}), with initial power spectra of the form 
(\ref{pwschi}) and uncorrelated random initial phases for $\chi_{\k}$, 
on $64^{3}$ cubic lattices with periodic boundary conditions. 
We have chosen the unit of length so that $k_{0}=2\pi$, and the unit of time
so that $2m = 1$. 
The integrations were done using a second-order in 
time operator-splitting algorithm; updates corresponding to the operator
$\nabla^{2}$ were carried out by a spectral method based on the fast Fourier
transform. The algorithm conserves the number of particles exactly;
energy nonconservation was below 1\% for all the data sets represented 
in the figures.

As we have mentioned in the introduction, there are basically two 
choices for the size of the integration box. One possibility is to choose 
the box large enough to activate collisionless instability, in order to see 
a crossover from the collisionless instability to a different regime. 
While it is clear, even a priori, that a linear instability cannot 
carry on forever and will have to end somehow, it is still 
interesting to see explicitly how that happens and what the different
regime is. Representative 
results are shown in Fig. 1. These results are for $B=20$ and box 
size $L=4.25$. A crossover, around $t=1.2$,
from a more rapid to a slower overall growth is
seen both in the density contrast and, especially, in the ratio of the 
total potential and kinetic energies. The density contrast is defined 
as
\be
\delta n / n = \langle  (\nu  - \nu_{\rm ave})^{2} \rangle^{1/2} / 
\nu_{\rm ave} \; ,
\label{contr}
\ee
where the brackets denote averaging over space. It is significant 
that the growth of the density contrast continues after the linear 
instability apparently terminates. This allows us to speak about a 
{\em nonlinear instability}, which we attribute to gravitational scattering.
The overall growth of the density contrast is superimposed on rapid 
oscillations, which are especially large at the later, scattering stage.
We attribute these oscillations in the density contrast to 
oscillations of matter about a forming star.

To confirm the existence of a nonlinear instability, it is useful to 
run simulations in a smaller box, so that the long-wave modes 
that could undergo the collisionless (linear) instability are eliminated.
For a cubic box with periodic boundary 
conditions, this means choosing the side length $L$ 
so that $2 \pi / L$ is larger than the wavenumber $k_{*}$ that appears 
on the right-hand side of (\ref{kcond})
(there are no density perturbations with $k=0$ because the total number of 
particles is conserved).
Any growth of the density contrast in an initially disordered gas in 
such a box will have to be attributed to some new type of gravitational 
instability. We can find out if this instability belongs to a new
``universality class''  by studying how its rate depends on the 
initial density of the gas. 

When gravity is the 
only interaction between the particles, the rate of clumping, during 
an initial stage when the gas is still nearly uniform,
can be written in general as
\be
t_{c}^{-1} = \frac{k_{0}^{2}}{2m} F(k_{J}^{4}/k_{0}^{4}, k_*L) \; .
\label{rate}
\ee
The function $F$ depends on two parameters that measure the 
strength of gravitational interaction between the particles:
\be
\xi = k_{J}^{4}/k_{0}^{4}
\label{xi}
\ee
measures the strength of interactions for momentum transfers
of order $2\pi/k_{0}$, while $k_{*} L$ measures that for momentum transfers
of order $k_{\rm min} = 2\pi / L$.
In the limit $k_{*} L \to \infty$, when the initial clumping 
is due to collisionless instability, $F$ becomes a function of its 
first argument only, and
\be
F(\xi) \propto \sqrt{\xi} \; ,
\label{F1}
\ee
see (\ref{disp2}). In an ordered Bose gas, dependence of the rate on $L$ disappears 
whenever $k_{J} \gg k_{\rm min}$, and it scaling with $\xi$ is then given by 
the same square-root law (\ref{F1}), see (\ref{disp1}).

In an initially disordered gas, situation is more complicated. The 
condition (\ref{disord}) (i.e. $\xi \ll 1$) guarantees only that interactions 
in the initial state are weak for momentum transfers of order $2\pi/k_{0}$. 
In a small box, we also have the condition $k_{*} < 
k_{\min}$, which guarantees that interactions are at most of moderate 
strength even for the smallest momentum transfers. In this case, the initial gas 
can be considered weakly interacting, and one may contemplate applying kinetic
theory. However, even if suitable kinetic equations exist,
they will, in general, not reduce to a Boltzmann equation (properly modified 
to include the effect of large occupation numbers). Indeed, as we have
already noted,
fluctuations with spatial scales $\ell \alt 2\pi/k_0$ cannot be described using the 
conventional semiclassical distribution
function. If such fluctuations are important (and we will argue that in the present
case they are), the Boltzmann equation will not apply.

The simplest way to see that the Boltzmann equation is in general
insufficient for studying evolution of a Bose gas, either for long-range or 
for short-range interactions, is to note that it includes interaction only via the scattering 
probability and so does not distinguish between attractive and repulsive interactions.
On the other hand, on physical grounds we expect dynamics for these two types of
interactions to be vastly different. 

Let us refer to the rate 
predicted by the Boltzmann equation for changes in the distribution 
function as the Boltzmann rate.
For a short-range attractive interaction, it has been found numerically in ref. 
\cite{dew} that scaling of the clumping rate with $\xi$ agrees to a good accuracy
with scaling of the Boltzmann rate. In general, however, there is no reason to expect 
such agreement. For gravitational scattering, the Boltzmann rate (for a gas with
large occupation numbers) is of the form (\ref{rate}) with
\be
F(\xi, k_* L) \propto \xi^{2} \ln(k_0 L) \; .
\label{FB}
\ee
The logarithm here is analogous to the Coulomb logarithm in plasma kinetics 
(cf. ref. \cite{LP});
it is due to an infrared divergence in the collision integral. In a small box,
this divergence is cut off at wavenumber transfers of order $k_{\min} = 2\pi /L$.
Now, consider the case when $k_0$ and $L$ are fixed. 
The scaling predicted by (\ref{FB}) is $F(\xi) \propto \xi^2$.
This is not quite what we see numerically. Instead, a good fit to the rate is obtained
with
\be
F(\xi) \propto \xi^{\alpha} \; .
\label{F2}
\ee
where $\alpha$ is noticeably larger than 2.

We attribute this large deviation of $\alpha$ from 2 to an important role
played by fluctuations with spatial scales much smaller than $2\pi/k_0$.
One can readily invent a reason why such fluctuations are more important in the presence 
of a long-range interactions than in the absence of such: even a short-scale 
fluctuation will influence many particles when there is a long-range force.
We thus arrive at a picture of short-scale fluctuations constantly appearing and 
disappearing, until one of them is able to start a growing clump. 

To extract the scaling law from our numerical results, we have used data corresponding 
to $L=3.5$ and three different values of $B$: $B=7$, $B=10$ and $B=20$. 
The corresponding values 
of $\xi=k_J^4/k_0^4$ are 0.050, 0.071, and 0.143. For these values of 
$L$ and $\xi$ ,
the right-hand side of (\ref{k*}) is smaller than $2\pi / L$, so collisionless
instability does not occur. Consequently, the density contrast's growth, which was
observed in all three cases, is a 
signature of a new type of gravitational instability. 

The growth of the density contrast should be more accurately referred 
to as a growth of some average with respect to time, because it is 
superimposed on rapid oscillations similar to those seen at later 
stages in Fig. 1. We expect that such an average will depend on time 
and $\xi$ mainly through a dependence on the ratio $t/t_{c}$, or, 
because $k_{0}$ and $L$ are kept constant, through the product 
$t\xi^{\alpha}$, where $\alpha$ is the power that we want to determine:
\be
\overline{\delta n / n} \approx f(t \xi^{\alpha}) \; .
\label{collapse}
\ee
The overline denotes averaging over several oscillations, and $f$ is 
some function.
When we plot the averaged density contrast,
for the above three values of $B$, against the rescaled time variable $t\xi^{\alpha}$, 
we find that a good collapse of the plots is achieved for
\be
\alpha = 2.7 \; ,
\label{alpha}
\ee
see Fig. 2.
The collapse is not nearly as good for $\alpha =2$, the value characteristic of
a Boltzmann rate. Our interpretation is that the instability is a result of gravitational
scattering, in which fluctuations with spatial scales $\ell \ll 2\pi/k_0$
play an important role.

There is an approximately linear stage in the growth of $\overline{\delta n/n}$
in Fig. 2, from about 0.5 to about 1.2 in units of $100 t \xi^{2.7}$.
Fitting this linear growth with a time dependence of the form
\be
\overline{\delta n / n} = t/t_{c} + {\rm const} \; ,
\label{fit}
\ee
provides one possible definition of the rate $t_{c}^{-1}$. 
Numerical fitting gives
\be
t_{c}^{-1} \approx 250 \xi^{2.7} \; .
\label{rate1}
\ee
This corresponds to eq. (\ref{rate}) with $F(\xi)\approx 6 \xi^{2.7}$.

The estimate (\ref{rate1}) has been 
obtained for initial power spectra of the form (\ref{pws}). It is 
natural to expect that a different form of the initial power 
spectrum will lead to a different numerical coefficient in (\ref{rate1}).
That is indeed so, although in cases that we have considered the 
difference is not overwhelming. For instance, one can generate a
random field with a power spectrum of the form
(\ref{pws}) and then make, by hand,
the magnitudes of $\chi$ on all lattice sites equal to some fixed 
value, while preserving the phases (in coordinate space). 
One can then use this new field 
as an initial condition that has a power spectrum different from
(\ref{pws}). We have done that for $L=3.5$, and $B=10$ and $20$. 
In this case, the density contrast is initially zero. However, it rapidly 
develops to a nonzero value, as dynamics of the density
is activated by dynamics of the phase of $\psi$.
The subsequent evolution of the density contrast is fairly similar to that
seen in Fig. 2. The scaling of the rate is consistent with $\alpha=2.7$,
but the rate itself is about half of that given by eq. (\ref{rate1}).

Finally, Fig. 3 shows the profile of the star at different moments of time,
and Fig. 4 illustrates the star's phase-correlated (Bose-Einstein condensed)
nature. (Only one large star has formed per box, in addition
to a number of much smaller clumps.)

\section{Discussion}
The main result of this work is numerical evidence for a new type of gravitational
instability. We have observed this instability in simulations of a disordered Bose gas 
in a box of a fixed size. 
We interpret it as being due to collisions between the particles, or, in 
terminology more suitable for a system with large occupation 
numbers, between classical waves. We have found that
scaling of the instability rate with the average density deviates from
the behavior characteristic of changes described the Boltzmann equation. We interpret
this deviation as an indicator of an important role played by 
short-scale fluctuations of the field. We have also presented evidence (Fig. 1) that this
instability will take over from a collisionless instability in a 
large system. This means that it could contribute to formation of structures
inside regions that had already fallen out of the Hubble expansion. In addition, we have
found that this collisional instability leads to formation of
a phase-correlated clump of matter (a Bose star) in an initially disordered Bose gas.
Hence, Bose stars can form through gravity alone even in
the absence of a large preexisting correlation length.

These results were obtained using the classical approximation for a Bose field.
The classical approximation applies to a field that
has a certain type of nonthermal power spectrum, namely, 
occupation numbers are large at small wavenumbers but rapidly 
decrease towards the ultraviolet. 
We do not exclude, however, that a similar nonlinear instability exists in
matter with different distributions of particles with respect to momenta, for example,
in a thermal or nearly thermal gas (for which the classical approximation does not apply).

Our results may apply to nonbaryonic dark matter (if the dark-matter particle is a boson)
and to low-density hydrogen. (At high densities, when inelastic 
processes are important,
the simple description (\ref{eqpsi}) is no longer adequate.) It is fascinating to think
that some of the matter in the universe may be in the form
of superfluid hydrogen. However, there may be interesting implications for structure 
formation even if our results apply only during initial stages of clumping, when
the density is still low.

We should also comment on a possible role of short-range interactions
in a disordered self-gravitating Bose gas.
As we have already mentioned, at low enough densities all the short-range 
interactions will be encoded in a single coefficient $g_{4}$ of the 
cubic term in eq. (\ref{eqpsi}). The strength of this local
interaction relative to the strength of gravitational scattering in 
the initial state is measured by the parameter 
$\eta=g_{4} k_{0}^{2}/4\pi G m^{2}$. This parameter can be small even when $g_{4}$
is much larger than $G/c^{2}$ ($c$ is the speed of light), provided 
the ratio $k_{0}/mc$ is small enough, i.e. the initial state is 
sufficiently nonrelativistic.
On the other hand, a not-too-small $\eta$ can lead to 
interesting effects. A repulsive interaction ($g_{4}>0$) in a Bose gas
causes coarsening---growth of the sizes of correlated domains 
\cite{coars}. If the interaction is strong enough, we expect that the gas will 
become sufficiently ordered before the nonlinear
instability had a chance to develop, and clumping will take 
place via the corresponding linear instability. 
In addition, we expect that repulsion will make emerging Bose stars larger
in size and smaller in density. (For equilibrium configurations, it 
has been long known that 
even a small repulsion makes a large difference \cite{Colpi&al}.) 
An attractive interaction ($g_{4}<0$)
produces correlated clumps all by itself \cite{dew}; we expect that it 
will help gravity along when both are present. 

Finally, we discuss relation of our work to earlier work in the 
literature. We can discern two trends in the earlier work on formation
of Bose stars. Some
researchers considered gravitational stability and evolution of 
already coherent configurations.
That work included linear stability analysis of a homogeneous 
Bose-Einstein condensate \cite{Khlopov&al,Bianchi&al}
and a study of  spherically symmetric collapse of an already coherent 
spherically symmetric configuration \cite{Seidel&Suen}.
It has been suggested \cite{Bianchi&al,Grasso} that the Jeans-type instability of 
a homogeneous Bose-Einstein condensate may lead to formation of 
Bose stars. Although we do not present corresponding data in this
paper, we have confirmed that a Bose star indeed forms in an ordered
Bose gas. However, we have given evidence that a large
preexisting correlation length, i.e. a high initial degree of spatial coherence, 
is not necessary for Bose star formation. A Bose star can form even
in an initially incoherent (disordered) gas via a new type of gravitational
instability that we have identified here.
The second trend in the earlier work was to consider ordering 
effects of local interactions.
In particular, it has been argued that due to effects of the Bose 
statistics even a very weak interaction, such as that between axions, 
can result in a relaxation time comparable to the age of the universe
\cite{Tkachev}, and, as a consequence, a Bose star may form out 
of an initially incoherent clump \cite{Kolb&Tkachev}. Here we have shown
that phase coherence can develop due to gravity alone, in the absence of any
additional interaction.

The author thanks F. Finelli and I. Tkachev for discussions. This work was supported 
in part by the U.S. Department of Energy under Grant DE-FG02-91ER40681 (Task B).

\appendix
\section*{More on the collisionless instability}
In this appendix we will show that for a wide class of initial 
distribution functions the collisionless instability (such as found in the Vlasov
approximation) disappears
at sufficiently large wavenumbers. Whether the instability is present or absent
is determined from the dispersion relation (\ref{dispeq}), which we rewrite 
here as
\be
\gamma(\omega, k) = 0 \; ,
\label{disp0}
\ee
where $\gamma$ is the ``gravitational permittivity'':
\be
\gamma(\omega, k) =
1 + \frac{4\pi G m^{3}}{k^2} \int_{-\infty}^{\infty} dp_x \frac{g'(p_x)}
{p_x - m\omega/k - i0} \; .
\label{gamma}
\ee
We have oriented coordinate axes so that $\k=(k, 0, 0)$, with $k>0$, 
and defined a distribution function with respect to $p_x$:
\be
g(p_x) = \int dp_y dp_z f_0(p_x, p_y, p_z) \; .
\label{gfunc}
\ee
In an isotropic medium, the form of $g(p_x)$ would not depend on the direction 
of $x$ (that is the direction of $\k$), but we do not assume isotropy here.

By its physical meaning, $g(p_{x})$ is nonnegative.
We assume that for any direction of $\k$ the function
$g(p_x)$ satisfies the following conditions: 
(i) $g(p_x)$ is smooth
(infinitely differentiable) and decreases with $|p_x|$ rapidly 
enough for the integral in (\ref{gamma}) to be convergent for any $\omega$ with
${\rm Im} \omega > 0$.
(ii) $g(p_x)$ is even; (iii) $g(p_x)$ is monotonically decreasing for all $p_x > 0$. From
these conditions, two more follow: (iv) $g'(0) = 0$; and (v) $g''(0) < 0$. Primes denote
derivatives with respect to $p_{x}$.

Eq. (\ref{gamma}) defines an analytic function in the upper half-plane of complex 
$\omega$; in our definition, the upper half-plane does not include the real axis. 
To obtain $\gamma(\omega, k)$ in the lower half-plane, we have to analytically 
continue it there from the upper half-plane. (Using eq. (\ref{gamma}) directly
in the lower half-plane would give another, unphysical sheet 
of $\gamma(\omega, k)$.)

Because $g(p_x)$ is even, $g'(p_x)$ is odd. From this, it follows that
in the upper half-plane $\gamma(\omega, k)$ can have 
zeroes only on the imaginary axis. For
$\omega = i \omega_{i}$ with $\omega_{i} > 0$, $\gamma$ takes the form
\be
\gamma(i\omega_{i}, k) =
1 + 4\pi G m^{3} \int_{-\infty}^{\infty} dp_x \frac{p_x g'(p_x)}
{k^{2} p_x^2 + m^{2} \omega_{i}^{2}} \; 
\label{gamma1}
\ee
(in particular, it is purely real). The conditions on $g(p_{x})$ 
guarantee that $p_{x} g'(p_{x}) < 0$ for any nonzero $p_{x}$. That 
means that the integral in (\ref{gamma1}) is negative, so that at any 
fixed $k>0$ $\gamma(i\omega_{i}, k)$ is a monotonically increasing 
function of $\omega_{i}$. Consequently, at a given $k$ it has at most one zero 
in the upper half-plane of $\omega$. This zero, when it occurs, signals 
an instability of the initial state with respect 
to fluctuations with that $k$. For example, 
such a zero always exists for $k\to 0$; its position, $\omega_{i}(0)$, is given 
by eq. (\ref{cless}). 

Now let us see what happens as $k$ increases.
Zeroes of $\gamma(\omega, k)$, i.e. roots of eq. (\ref{disp0}), define a dispersion law
$\omega(k)$. As long as, for a given $k$, the zero is still in the upper half-plane, we 
can continue to use eq. (\ref{gamma1}). First, it follows from 
(\ref{gamma1}) that $\omega_{i}(0)$ is 
an upper bound on $\omega_{i}(k)$. So, as we change $k$, a root cannot 
appear from infinity. Second, differentiating 
(\ref{disp0}) with respect to $k$ and using again (\ref{gamma1}), 
we find that, when $g(p_{x})$ obeys 
the stated conditions, $d\omega_{i}(k)/dk$ is necessarily negative. 
So, as $k$ increases, the root moves down towards the real axis. It 
reaches the real axis at the value $k=k_{*}$ determined by setting 
$\omega_{i}=0$ in (\ref{gamma1}):
\be
k_*^2 = 4\pi G m^{3} \left| \int_{-\infty}^{\infty} dp_x g'(p_x) / p_x 
\right| \; .
\label{k*1}
\ee
(Note that there is no divergence at $p_x=0$ here, because $g'(0)=0$, 
see condition (iv).) We have already established that
any root of (\ref{disp0}) in the upper half-plane is on the 
imaginary axis, and that such a root cannot appear from infinity. As a 
consequence, the point $\omega = 0$ is the only
point where the root of (\ref{disp0}) can leave or enter the upper half-plane
of $\omega$. Because the corresponding value of $k$ is unique, the root will not
reappear in the upper half-plane at any $k > k_{*}$.
We conclude that the collisionless instability is absent for all $k\geq k_{*}$.

A detailed picture of how the instability disappears can be obtained
by expanding the original expression for $\gamma$, eq. (\ref{gamma}), in an
asymptotic series about $\omega=0$. The $i0$ prescription in (\ref{gamma}) will
ensure that we are expanding on the correct sheet of $\gamma(\omega, k)$:
\be
\gamma(\omega, k) =
1 + \frac{4\pi G m^{3}}{k^2} \int_{-\infty}^{\infty} dp_x \frac{g'(p_x)}
{p_x  - i0} \left( 1 + \frac{m \omega / k}{p_x  - i0}  + O(\omega^2) \right) \; .
\label{gamma2}
\ee
For a smooth even $g(p_x)$, this evaluates to
\be
\gamma(\omega, k) =
1 - \frac{k_*^2}{k^2} \left( 1 + 
\frac{i \pi \omega m g''(0) / k}
{\int_{-\infty}^{\infty} dp_x g'(p_x) / p_x} + O(\omega^2) \right) \; .
\label{gamma3}
\ee
Writing $k=k_* + \Delta k$ and expanding in small $\Delta k$, we find that the root
of the dispersion equation has the form
\be
\omega(\Delta k) = - i \frac{2\Delta k}{\pi m} \left| 
\frac{\int_{-\infty}^{\infty} dp_x g'(p_x) / p_x}{g''(0)} \right| 
+ O((\Delta k)^2) \; .
\label{ome_sol}
\ee
This expression shows explicitly that, as $k$ increases past $k_*$, the root 
moves from the upper half-plane into the lower half-plane.

The structure of higher-order terms in (\ref{gamma3}) is as follows: the coefficients
of even powers of $\omega$ are all real, while the coefficients of odd powers 
are all imaginary.  So, to any finite order in $\Delta k$, the root of the
dispersion relation will remain purely imaginary even after it moves in the lower 
half-plane, and the mode corresponding to that $\Delta k$ will
be overdamped (non-oscillatory). This conclusion may be obviated by terms that are
``nonperturbative'' in $\Delta k$, i.e. are not seen in the asymptotic expansion
(\ref{gamma2}). We also stress that it applies only to fluctuations about 
a homogeneous state, not to a state that already contains clumps.

\begin{figure}
\leavevmode\epsfysize=3in \epsfbox{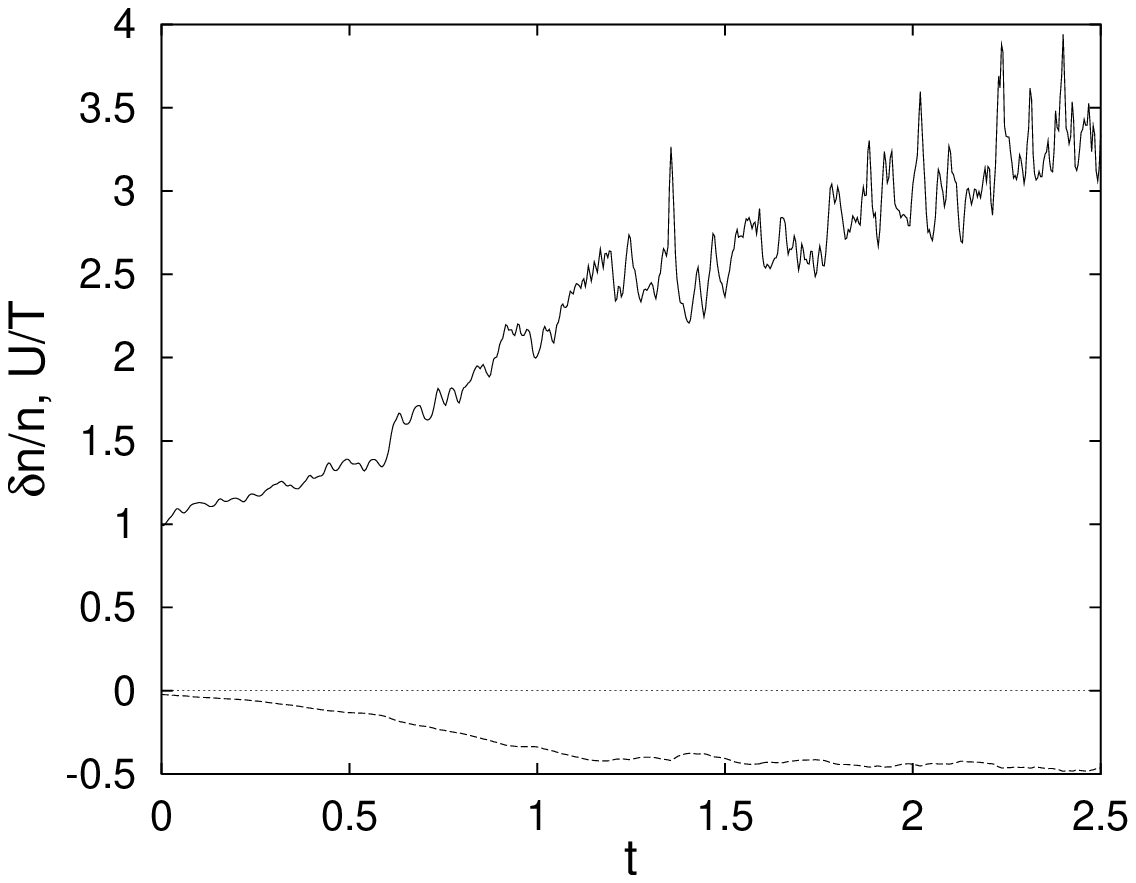}
\vspace*{0.2in}
\caption{Illustrates crossover from the collisionless instability to a scattering
regime. These plots are for $B=20$ and $L=4.25$. Solid line shows the density contrast;
dashed line---the ratio of the total potential and kinetic energies. 
The crossover occurs around $t=1.2$.
}
\label{fig:crossover}
\end{figure}

\begin{figure}
\leavevmode\epsfysize=3in \epsfbox{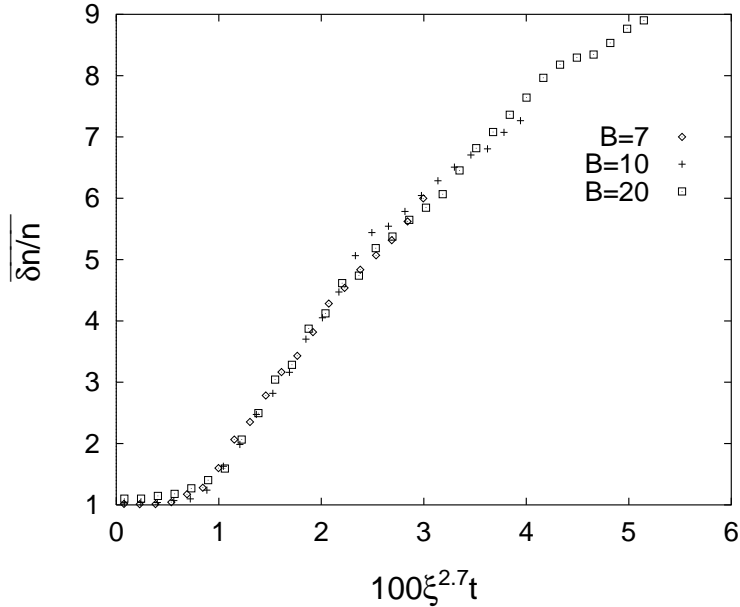}
\vspace*{0.2in}
\caption{Density contrast averaged over intervals of time, as a function of 
rescaled time. Each symbol is at the center of an interval over which 
the averaging is carried out. All data are for $L=3.5$. The data for $B=7$ and $B=20$
correspond to the same realization of random initial phases; those for 
$B=10$---to a different one.
}
\label{fig:collapse}
\end{figure}

\begin{figure}
\leavevmode\epsfysize=3in \epsfbox{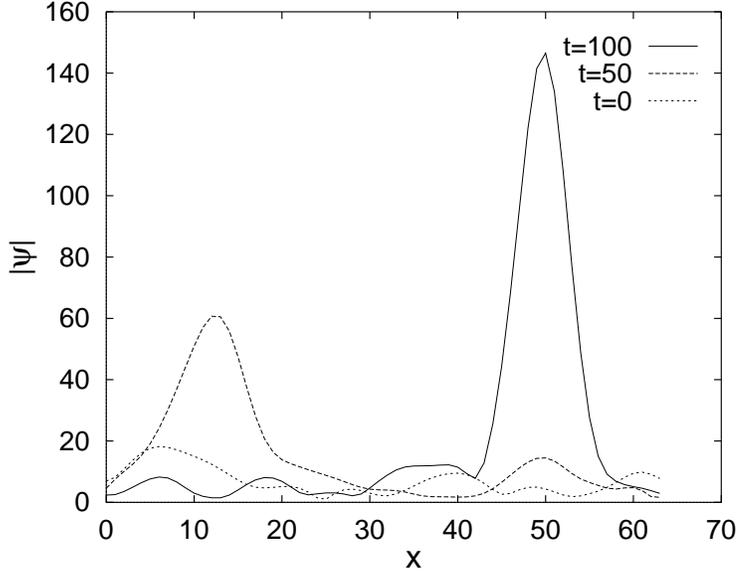}
\vspace*{0.2in}
\caption{Profile of $|\psi|$ along a line passing through the maximum of $|\psi|$,
at different moment of time. Because the location of the maximum moves, the actual
lines are different at different moments (although they are chosen parallel).
These plots are for $L=3.5$ and $B=7$.
The coordinate along the line is in units of the lattice spacing.
}
\label{fig:profiles}
\end{figure}

\begin{figure}
\leavevmode\epsfysize=3in \epsfbox{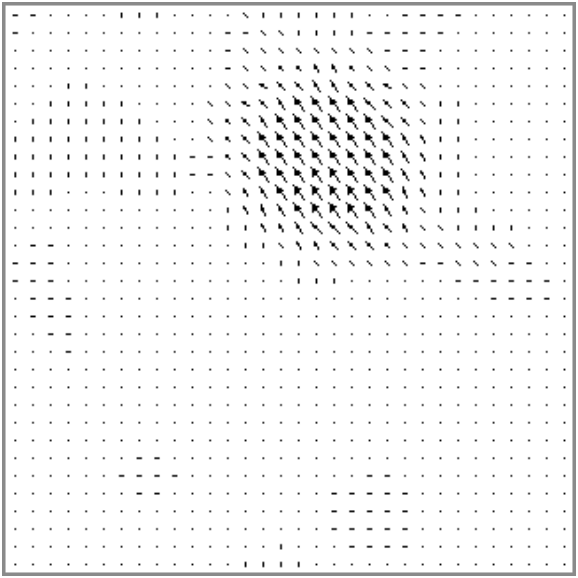}
\vspace*{0.2in}
\caption{
Distribution of the field $\psi$ on a quarter of a spatial slice at time $t=100$, 
for $L=3.5$ and $B=7$. The slice cuts through the star's center.
The length of an arrow corresponds to $|\psi|$ on a given site, and the angle the arrow 
makes clockwise from 12 noon---to the phase of $\psi$. Arrows on all sites with 
$|\psi| \geq 70$ have the same (maximal) length.
}
\label{fig:slice}
\end{figure}

\end{document}